\documentclass[reprint,superscriptaddress,amsmath,amssymb,aps,pra]{revtex4-2}
\usepackage{graphicx}                                                                                                                                                                                                                                                                                                       
\usepackage{dcolumn}
\usepackage{bm}

\begin{document}

\preprint{AIP/123-QED}

\title{One-Dimensional Quench Dynamics in an Optical Lattice: sine-Gordon and Bose-Hubbard Descriptions}

\author{Subhrajyoti Roy}
\affiliation{Department of Physics, Presidency University, 86/1   College Street, Kolkata 700073, India.}

\author{Rhombik Roy}
\thanks{Currently at: Department of Physics, University of Haifa, 3498838 Haifa, Israel.}
\affiliation{Department of Physics, Presidency University, 86/1   College Street, Kolkata 700073, India.}

\author{Andrea Trombettoni}
\affiliation{Department of Physics, University of Trieste, Strada Costiera 11, I-34151 Trieste, Italy}
\affiliation{CNR-IOM DEMOCRITOS Simulation Centre and SISSA, Via Bonomea 265, I-34136 Trieste, Italy.}

\author{Barnali Chakrabarti}
\email{barnali.physics@presiuniv.ac.in}
\affiliation{Department of Physics, Presidency University, 86/1   College Street, Kolkata 700073, India.}
\affiliation{The Abdus Salam International Center for Theoretical Physics, 34100 Trieste, Italy.}
\affiliation{Instituto de Física, Universidade de São Paulo, CEP 05508-090, SP, Brazil.}

\author{Arnaldo Gammal}
\affiliation{Instituto de Física, Universidade de São Paulo, CEP 05508-090, SP, Brazil.}


\begin{abstract}
We investigate the dynamics of one-dimensional interacting bosons in an optical lattice after a sudden quench in the Bose-Hubbard (BH) and sine-Gordon (SG) regimes. While in higher dimension, the Mott-superfluid phase transition is observed for weakly interacting bosons in deep lattices, in 1D an instability is generated also for shallow lattices with a commensurate periodic potential pinning the atoms to the Mott state through a transition described by the SG model. The present work aims at identifying the SG and BH regimes. We study them by dynamical measures of several key quantities. We numerically exactly solve the time dependent Schr\"odinger equation for small number of atoms and investigate the corresponding quantum many-body dynamics. In both cases, correlation dynamics exhibits collapse revival phenomena, though with different time scales. We argue that the dynamical fragmentation is a convenient quantity to distinguish the dynamics specially near the pinning zone. To understand the relaxation process we measure the many-body information entropy. BH dynamics clearly establishes the possible relaxation to the maximum entropy state determined by the Gaussian orthogonal ensemble of random matrices (GOE). In contrast, the SG dynamics is so fast that it does not exhibit any signature of relaxation in the present time scale of computation. 
\end{abstract}


\maketitle

\section{Introduction} \label{sec:intro}

Ultracold atomic gases in optical lattices offer a well known laboratory system with unprecedented control over experimental parameters. It has emerged in the last two decades as an ideal test bed for the study of complex many-body quantum phenomena~\cite{RevModPhys.80.885}. Experimentally it became quite straightforward to tune the lattice depth of the optical potential and study different dynamical regimes and physical phenomena, such as quantum phase transitions~\cite{nature.415} and Josephson dynamics \cite{cataliotti01}, see more references in \cite{Bloch2005,oberthaler,book1,Gross17,Bloch2022}.

Quantum many-body systems may exhibit quantum phase transitions at zero temperature, where thermal fluctuations are frozen out, but strong quantum fluctuations drive the system from one phase to another \cite{sachdev}. Across the phase transition two competing energy terms in the many-body Hamiltonian may determine the presence of a critical value in the ratio of the system parameters. A major example of quantum phase transition is the one from a superfluid (SF) phase to Mott insulator phase (MI)--also referred to as the Mott-Hubbard quantum phase transition~\cite{nature.415,science.1192368}. The Mott transition in dimension larger than $1$ takes place for interacting bosons in deep lattices, and it is driven by the competition between the interaction strength ($U$) and hopping term ($J$). In the experiments reported in ~\cite{nature.415,nature.419}, Rb atoms loaded in optical lattices demonstrated phase transition to insulating Mott phase at a critical value $J/U$~\cite{nature.419} and a periodic series of collapse and revival in the dynamical evolution of the matter-wave interference pattern was observed~\cite{nature.419}.

On the theoretical side, it is well known that the lattice model-Bose Hubbard (BH) captures the relevant physics and the SF to MI transition is well studied in all dimensions~\cite{book1}. In the SF phase, the effective dynamics is described by the classical version of the BH model, that is the discrete nonlinear Schr\"odinger equation~\cite{trombettoni01,review_panos}. The study of the properties of strongly interacting bosons confined in one-dimension is particularly interesting as the effects of quantum fluctuations and correlations are enhanced. Especially, the extreme case of the Tonks-Girardeau 
gas when the strongly correlated bosons minimize their spatial overlap and acquire fermionic properties~\cite{science.305,haller}. In the limit of vanishing periodic potential, a remarkable quantum phase transition is observed--pinning the system by an arbitrarily weak lattice leads to sine-Gordon (SG) phase transition from SF Luttinger liquid to Mott insulator~\cite{Buchler}. In a shallow potential, the MI transition is basically controlled by the strong interaction. Experimentally this Mott transition has been observed through modulation spectroscopy and transport phenomena ~\cite{Haller:2010}. The pinning transition in the vanishing lattice is described by (1+1) sine-Gordon model~\cite{Coleman,Mussardo}.

Despite of the existence of outnumbering research work to understand the Bose-Hubbard physics in deep lattices~\cite{fischer,Bloch2005,oberthaler,book1,Gross17,Jaksch:1998,RevModPhys.80.885,Esslinger:2004}, the existing literature are relatively less developed for strongly interacting bosons in shallow lattices. The latter is a challenge as the BH model is restricted to cases in which one can define site localized Wannier functions~\cite{Astrakharchik:2016}. To address strongly correlated bosons in shallow lattices, it is necessary to go beyond the BH model~\cite{Sascha:2010}. Mott transition for strongly interacting bosons in shallow lattices was studied in ~\cite{Boeris,Astrakharchik:2016} using quantum Monte Carlo simulations in the continuum space. The main motivation was to precisely determine the phase diagram in plane $\gamma$-$V_0$, where $\gamma$ is the dimensionless Lieb-Liniger coupling constant \cite{cazalilla2011,kormos09}, proportional to the strength of the two-body $\delta$ interparticle potential \cite{olshanii99} and $V_0$ is the strength of the periodic potential. It was confirmed that the line separating the SF and the MI phases is giving a finite value of $\gamma$ in the limit $V_0 \to 0$ for commensurate filling. It was also found a significant deviation from the perturbative sine-Gordon theory, and the regimes of validity of BH and SG models were also discussed ~\cite{Boeris,Astrakharchik:2016}. We finally mention that the effects of deviations from commensurate filings were also studied \cite{buchler03,lazarides09}.

In this paper, we address the two experimental situations:
a) pinning transition in shallow lattices~\cite{Haller:2010}. b) the collapse revival dynamics in deep lattice~\cite{nature.415}. The quench dynamics of strongly interacting superfluid in the shallow lattice is termed as SG dynamics and the same for weakly interacting superfluid in deep lattice is termed as BH dynamics. The main motivation is to clarify the intermediate correlated dynamics both in the SG and the BH regimes. The time evolution of the many-body Schr\"odinger equation is computed by the {\it ab initio} many-body method MCTDHB (multiconfigurational time dependent Hartree for bosons)~\cite{Streltsov:2006,Streltsov:2007,Alon:2007,Alon:2008,Lode:2016,Fasshauer:2016,Lode:2020}, which calculates the many-body wave function numerically exactly. 
We observe that the system exhibits {\it {rich and distinguishable}} many-body dynamics at different time scales in both regimes. We present and discuss the one- and two-body correlation functions and prescribe some key quantities to understand the difference of SG and BH dynamics. We find that the dynamical fragmentation plays the key role to distinguish SG dynamics from BH dynamics, especially in the pinning zone of sine-Gordon regime. We establish that the time required for first entry in the Mott state, natural occupation, Shannon information entropy and the local density-density correlation are the important measures to distinguish the SG and BH dynamics.  

For the SG regime, where the initial state is fragmented and strongly correlated due to strong interatomic interaction, upon quench it enters fast to the Mott phase. Whereas in the BH regime, the initial state is a non-fragmented and less correlated superfluid, upon quench it enters the Mott phase at a time which is significantly larger than the time scale in SG regime. We numerically calculate many-body information entropy in the long-time dynamics to understand the possibility of statistical relaxation in both cases. For the BH dynamics, after the initial SF-MI collapse-revival, the system exhibits some modulated oscillations in entropy, presenting signatures of a possible relaxation at much longer time. At variance, for the SG dynamics we observe unmodulated collapse-revival phenomena in the maximum time period achieved in the present set of computation (of course, it may eventually relax at a very long time which is out of scope of the available simulation).

The paper is organized as follows. In Sec.\ref{sec:II}, we introduce the theoretical framework and the quantities of interest.  Sec.\ref{sec:III} presents the initial set up and quench protocol. Sec.\ref{sec:IV} presents the first-order Glauber correlation function both in the SG and BH dynamics. Sec.\ref{sec:V} presents the second-order Glauber correlation function. Sec.\ref{sec:VI} deals with the several measures used to distinguish the SG dynamics from the BH dynamics.  Sec.\ref{sec:VII} concludes the summary. Appendix A discusses the main results for larger system size and Appendix B summarizes the units used in the numerical computation. 

\section{The Model}
\label{sec:II}

We consider $N$ interacting bosons, each of mass $m$ with a contact interaction in a 1D periodic potential $V_{OL}(x)=V \sin^2(kx)$, 
with $V$ the lattice depth and $k$  the wave vector.
The characteristic energy associated with the lattice is the recoil energy $E_r=\hbar^2k^2/2m$.
The full many-body Hamiltonian reads
\begin{equation}
    \hat{H}=\sum^N_{i=1}\left[ -\frac{\hbar^2}{2m}\frac{\partial^2}{{\partial x_i}^2}+V_{OL}(x_i)\right]+\lambda\sum_{i<j}\delta(x_i-x_j)
\label{ham}
\end{equation}
$\lambda$ is the interaction strength and is determined by the scattering length $a_0$ and the transverse confinement. For the present calculation we consider only repulsive interaction ($\lambda >0$). The Hamiltonian $H$ is scaled in terms of $E_r$, with $\hbar=m=k=1$, thus rendering all terms dimensionless. Please refer to the Appendix B for a complete discussion about the dimensionless unit used in the MCTDHX simulation. 

We solve the time dependent Schr\"odinger equation (TDSE)  $\hat{H} \psi = i \frac{\partial \psi}{\partial t}$ by using the (MCTDHB) ~\cite{Streltsov:2006,Streltsov:2007,Alon:2007,Alon:2008,Lode:2016,Fasshauer:2016,Lode:2020} with periodic boundary condition as implemented in MCTDH-X package ~\cite{Lin:2020,MCTDHX}. The MCTDHB method is a bosonic variant within the broader family of MCTDH methodologies~\cite{10.1063/1.4821350,10.1063/1.2902982,BECK20001,10.1063/1.1580111,Wang2015,PhysRevA.91.012509,variational4,Lévêque_2017}. The time-dependent many-body wave function is expanded in time-dependent permanents as $\vert\psi(t)\rangle=\sum_{\bar{n}}C_{\bar{n}}(t) \vert\bar{n},t\rangle$ where $\vert\bar{n},t\rangle=\vert n_1,n_2,\hdots,n_M;t \rangle$ represents the occupation numbers. The time-dependent single particle orbital $\{\phi_k(x,t)\}$ and the expansion coefficients $\{C_{\bar{n}}(t)\}$ are determined by the time-dependent variational principle~\cite{variational1,variational2,variational3,variational4}. Compared to a time-independent basis, as the permanents are time-dependent, a given degree of accuracy can be reached with much shorter expansion~\cite{mctdhb_exp1,mctdhb_exp2}. In MCTDHB, the time adaptive many-body basis set can dynamically follow the building correlation due to the inter-particle interaction~\cite{Alon:2008,Alon:2007, mctdhb_exact3,barnali_axel,rhombik_jpb,rhombik_pra,rhombik_quantumreports,rhombik_pre,rhombik_scipost,PhysRevA.109.063308}. 

From the many-body wave function $|\psi(t)\rangle$ we calculate several observables. To get the information about the spatial distribution of the bosons we calculate the one-body density as 
\begin{equation}
\rho(x;t)=\langle\psi(t)|\hat{\psi^\dag}(x)\hat{\psi}(x)|\psi(t)\rangle.
\end{equation}
To measure the degree of coherence in the many-body correlation dynamics we calculate the reduced one-body and two-body densities defined as 
\begin{equation}
\rho^{(1)}(x|x';t)=\langle\psi(t)|\hat{\psi^\dag}(x)\hat{\psi}(x')|\psi(t)\rangle
\end{equation}
\begin{equation}
\rho^{(2)}(x|x';t)=\langle\psi(t)|\hat{\psi^\dag}(x) \hat{\psi^\dag}(x') \hat{\psi}(x') \hat{\psi}(x)|\psi(t)\rangle
\end{equation}
We further calculate the one-body and two-body Glauber correlation functions as 
\begin{equation}
g^{(1)}(x, x';t) = \frac{\rho^{(1)}(x,x';t)}{N \sqrt{\rho(x) \rho(x')}} 
    \end{equation}
    
    \begin{equation}
     g^{(2)}(x, x';t) = \frac{\rho^{(2)}(x,x';t)}{N^2 \sqrt{\rho(x) \rho(x')}}  
\end{equation}
The diagonal elements of $g^{(p)}(x_{1}^{\prime}, \dots, x_{p}^{\prime}, x_{1}, \dots, x_{p};t)$, denoted as $g^{(p)}( x_{1}, \dots, x_{p};t)$, provide a measure of $p$-th order coherence. Here we restrict to $p=1$ and $2$ to determine the one- and two-body coherence in the dynamics. 
If $\vert g^{(p)}( x_{1}, \dots, x_{p};t) \vert = 1$, the system is fully coherent, while deviations from unity indicate partial coherence. Specifically, $g^{(p)}( x_{1}, \dots, x_{p};t) > 1$ implies correlated detection probabilities at positions $x_{1}, \dots, x_{p}$, while $g^{(p)}( x_{1}, \dots, x_{p};t) < 1$ indicates anti-correlations.\\

The information theoretical measure known as Shannon entropy is the ideal quantity to study the statistical relxation process in the quench dynamics. However the standard definition of Shannon information entropy is based on the real-space and momentum space one-body density which are insuffient to follow the built up correlation during time evolution~\cite{barnali_axel}. Here we define an equivalent measure of information entropy called occupation entropy as 
\begin{equation}
S_{info}(t) = -\sum_{n_i} \frac{n_i(t)}{N} \ln \left( \frac{n_i(t)}{N} \right),
\label{eq:S_n}
\end{equation}
where $n_i(t)$ is the occupation of the $i$-th element of the many-body configuration $\bar{n}$. In the Gross-Pitaevskii single orbital mean-field theory, as the reduced density matrix has a single eigenvalue, $S_{info}(t)$ is zero. However for fragmented many-body state several configurations may contribute, $S_{info} \neq 0$. As more and more orbitals start to populate the many-body state, $S_{info}(t)$ gradually increases and finally saturates to a value determined by the maximally fragmented many body state. For time reversal and rotationally invariant systems, the maximum entropy is also determined by the predictions of a Gaussian orthogonal ensemble of random matrices(GOE)~\cite{Kota,Rigol,Izrailev:2012,Mark:1994}. Importantly we also demonstrate that $S_{info}(t)$ approach the GOE value.

\section{Initial states and quench protocol}
\label{sec:III}
\begin{figure}
    \centering
    \includegraphics[width= 0.45\textwidth]{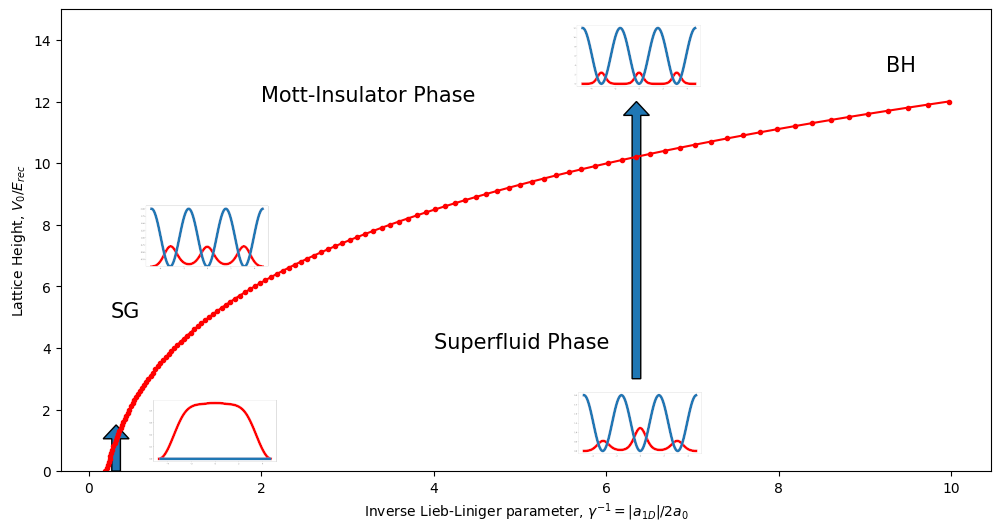}
    \caption{The zero-temperature phase diagram of the continuous model as a function of inverse of Lieb-Liniger parameter ($s$-wave scattering length) and the optical lattice depth. The red solid line is obtained from the exact diagonalization of Bose-Hubbard model. The data points are taken from the Ref~\cite{Astrakharchik:2016}. The red line separates the superfluid phase from the Mott phase. The arrow on the left hand side represents the quench protocol in the vanishing lattice which is termed as sine-Gordon dynamics. The arrow on the right side represents the quench protocol in the Bose-Hubbard (BH) regime which is termed as BH dynamics. In both cases the prequench and post quench states are inserted  for better guidance. All quantities shown are dimensionless; refer to the text for further details. }
    \label{fig:phase}
\end{figure}

With the Hamiltonian given in Eq. (\ref{ham}), we will study two different quench protocols, as summarized in Fig.1. 
In Fig.~\ref{fig:phase}, we report the phase diagram of the model taken from the Ref.~\cite{Astrakharchik:2016} for one particle per well. The solid red line is the transition line which separates the superfluid phase and the Mott insulator phase. The data points are  from the exact digonalization of Bose-Hubbard model~\cite{Astrakharchik:2016}. We utilize two arrows to specify the two different kinds of quench protocol addressed in this work. As described in the introduction, the quench in the strong interaction and shallow lattice region is termed as sine-Gordon dynamics, whereas the quench in the weak interaction and deep lattice is termed as Bose-Hubbard dynamics. The Lieb-Liniger parameter ($\gamma$) is related to the two-body interaction strength parameter ($\lambda) $ by $\gamma$ = $\frac{m \lambda}{\hbar^2 n} $, where $n= \frac{N}{L}$ is the density of the gas and $L$ is the length of the system. The critical value of Lieb Liniger parameter at the vanishing lattice $\gamma_c \approx 3.65$. Notice that in Ref.~\cite{Astrakharchik:2016} the phase diagram is extrapolated for an infinite system, while we start our investigation here for a system which is a small finite ensemble of three bosons in three lattice sites. We investigate the triple well system because it is the elemental building block of larger system that exhibits all essential dynamical features to distinguish SG and BH dynamics. In the Appendix A, we present key measures in the relaxation process for larger systems and demonstrate that the choice of our finite system does not dictate the physics.

For the SG quench dynamics we prepare the initial state in the vanishing lattice with $\frac{1}{\gamma}$ = 0.315 which corresponds to $\lambda=1.01$. Thus the prequench state is a strongly interacting superfluid in the vanishing lattice. Then at $t=0$, we propagate this state by pinning with a weak lattice of strength $V=1.5$ which leads to Mott localization. Thus our choice mimics the pinning transition of Ref.~\cite{Haller:2010} which is represented by a short arrow on the left hand side to mark the quench in sine-Gordon regime. The prequench and the post quench states are also presented schematically at the bottom and top of the arrow respectively. In absence of perturbation, the system is strongly correlated superfluid (bottom) and a weak perturbation induces transition to the insulating Mott state (top). 

Whereas the tall arrow on the right hand side marks the quench dynamics in the Bose-Hubbard region. We also presents the schematic density distributions. We prepare the initial state in a lattice of depth $V=3.0$ and $\frac{1}{\gamma}=6.36$ which corresponds to $\lambda=0.05$. Thus the prequench state is a superfluid phase (bottom). At $t=0$, we suddenly quench to $V=12.0$ which leads to Mott localization (top). Thus our choice of parameters mimics the SF to Mott transition of Greiner experiment~\cite{nature.419}. 

\section{First-order Glauber correlation dynamics}
\label{sec:IV}

In this section, we present the results for the First-order Glauber correlation function using the Eq.(5). We find that the dynamics of first-order coherence strongly depends on the initial correlation of the prequench state and the degree of fragmentation. For the prequench states in both regimes we prepare the system at $t=0$ in the ground state of lattice potential with suitable $V$ and interaction strength as described in Sec. \ref{sec:III}.  We have three bosons in three lattice sites, we impose periodic boundary condition. Then quench to appropriate $V$ chosen according to the sine-Gordon or Bose-Hubbard regime as shown in Fig.1. All throughout the dynamics we keep $M=6$ orbitals. We have probed convergence both in the initial state propagation and repeating the dynamics with $M=8$. We confirm that the dynamics is not affected by the higher number of orbitals.  

\begin{figure}
    \centering
    \includegraphics[width= 0.45\textwidth]{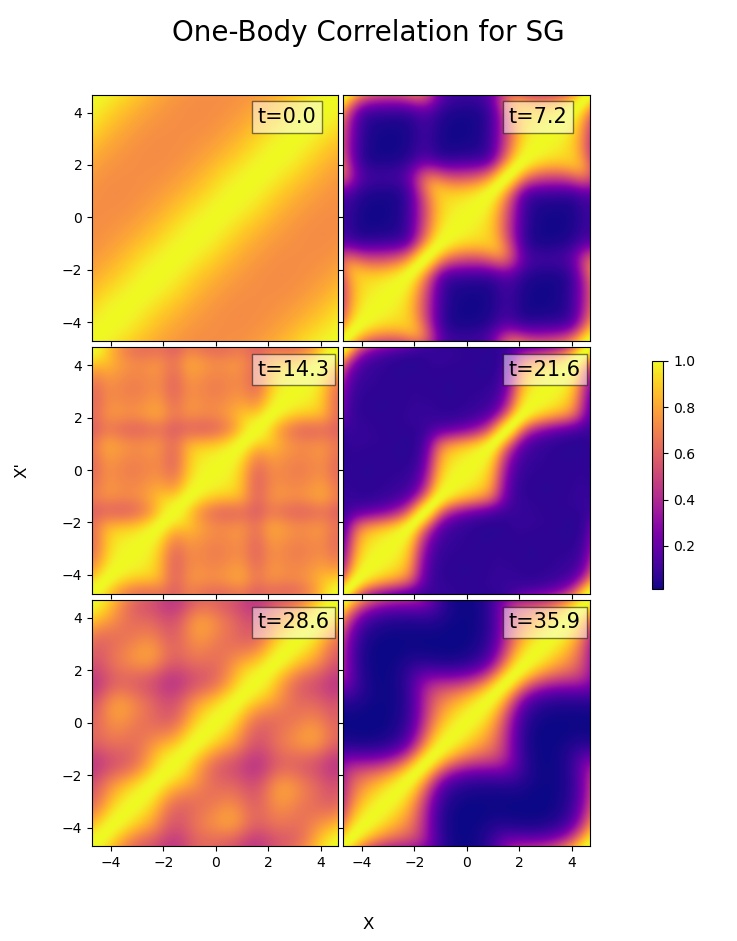}
    \caption{First-order Glauber correlation ($g^{(1)}(x,x';t)$) at different times for sine-Gordon transition. The computation is done with $N=3$ bosons in $S=3$ lattice sites and orbital $M=6$. The strongly interacting bosons are prepared in zero lattice and is suddenly pinned with a very weak lattice. The time for first Mott entry is $\approx$ 7.2. It reveals oscillations between the fragmented superfluid and Mott-insulator phases with a period of close to $t= 7.0$. All quantities shown are dimensionless; refer to the text for further details. }
    \label{fig:sg_1B}
\end{figure}
\begin{figure}
    \centering
    \includegraphics[width= 0.45\textwidth]{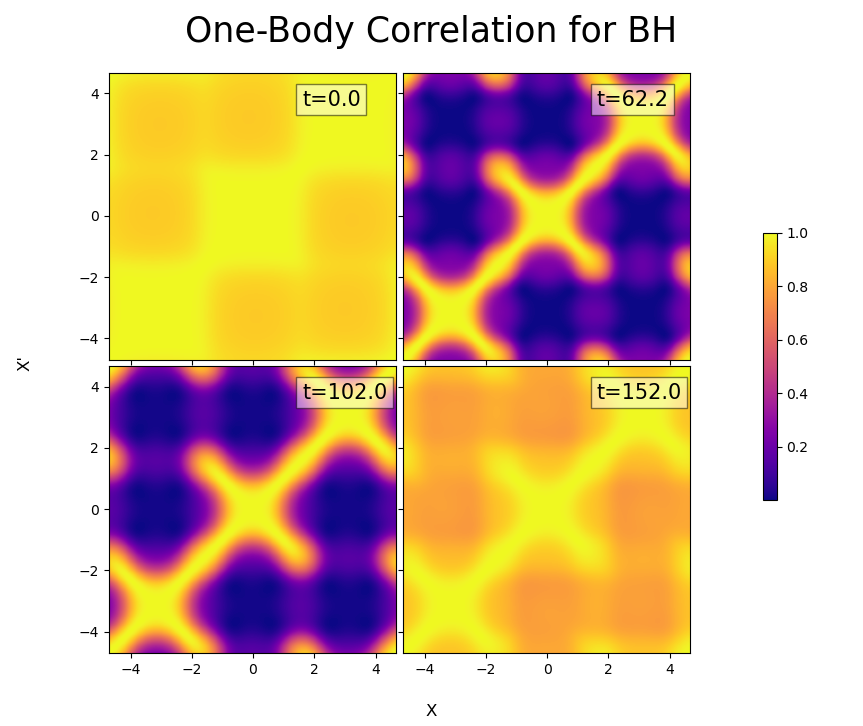}
    \caption{First-order Glauber correlation ($g^{(1)}(x,x';t)$) at different times for Bose-Hubdard transition. The computation is done with $N=3$ bosons in $S=3$ lattice sites and orbital $M=6$. Weakly interacting bosons in a weak lattice is suddenly quenched to a deep lattice in BH regime. The time required for first Mott entry is $\approx$ 62.2. Collapse-revival between the superfluid and Mott-insulator phases happens on a large time scale and of no fixed period. All quantities shown are dimensionless; refer to the text for further details. }
    \label{fig:bh_1B}
\end{figure}

\subsection{Correlation dynamics in the sine-Gordon regime} 
The initial state is prepared in the vanishing lattice with $\lambda= 1.01$ and suddenly pinned with $V=1.5$.
In Fig~\ref{fig:sg_1B} we show the results for one-body correlation function $g^{(1)}(x,x';t)$, snapshots are presented at different times. At $t=0$, the structureless bright diagonal exhibits strong diagonal correlation. The corresponding off-diagonal correlation is $\approx$ 70 \% compared to the scale of diagonal correlation $\approx$ 100 \%. This complex many-body correlation exhibits some many-body phase intermediate to the SF and MI phases. It is not a SF phase as the global correlation, $g^{(1)}$ $\approx$ 1.0,  across the lattice is not maintained. This is not a MI phase as the off-diagonal correlation is not extinguished, indeed perfect Mott localization does not happen. This intermediate phase arises due to the strong interplay between strong interaction and interatomic correlation as the lattice is subrelevent. The strong repulsive interaction pushes the on-site correlation outside the lattice, off-diagonal correlation is built up at the cost of diagonal correlation. Later, we will establish that this many-body state is a fragmented superfluid which is contributed by several natural orbitals.  At time $t = 7.2$, it enters what we can call the first Mott state, with three distinct lobes along the diagonal and absence of the off-diagonal correlation. We choose $t=7.2$ as the time for {\it{first Mott entry}} ($t_{entry}^{M1}$). We further follow the dynamics and observe periodic series of collapse and revival between fragmented superfluid and Mott state with average time period of $ \approx 7.0$. 

\subsection{Correlation dynamics in Bose-Hubbard regime}
The initial state is prepared in a weak lattice of depth $V=3.0$ and weak interaction with $\lambda=0.05$. It is suddenly quenched to deep lattice of depth $V=12.0$. The corresponding correlation dynamics as presented in Fig.~\ref{fig:bh_1B}, it nicely demonstrates the collapse revival dynamics but at a very large time scale. The initial state ($t=0$) is a pure superfluid state when diagonal to off-diagonal correlation varies between $1.0$ to $0.97$. The structure along the diagonal exhibits the presence of lattice, which was absent in Fig~\ref{fig:sg_1B} due to vanishing lattice. The time required for first Mott entry is quite large $t_{entry}^{M1} = 62.2$, the diagonal correlation is maintained, with extinction of off-diagonal correlation till the time $t=102.0$. This is the Mott holding time when some redistribution of correlation across the lattice occurs. At time $t=152.0$, the revival to superfluid occurs. There is no fixed period of collapse revival dynamics in the next cycle (not shown here) as observed in sine-Gordon dynamics. The time scale of dynamics is significantly larger for Bose-Hubbard dynamics. 

The difference in the time scale for BH and SG dynamics can be interpreted as follows. Both quench protocol leads to Mott localization and for an infinite system two Mott regimes are smoothly connected. However the quench dynamics also significantly depend on the initial correlation. In the sine-Gordon regime, Mott localization happens due to initial strong correlation raised from strong interaction, lattice is irrelevant. In the Bose-Hubbard regime, Mott localization happens due to dominating effect of lattice depth; the effect of weak interaction is irrelevant. Thus the Mott localizations in the post quench states are differed by their Mott gap which results to observe different time scale of dynamics in the localization process. 
\section{Second-order Glauber correlation dynamics}
\label{sec:V}
\begin{figure}
    \centering
    \includegraphics[width=0.5\textwidth]{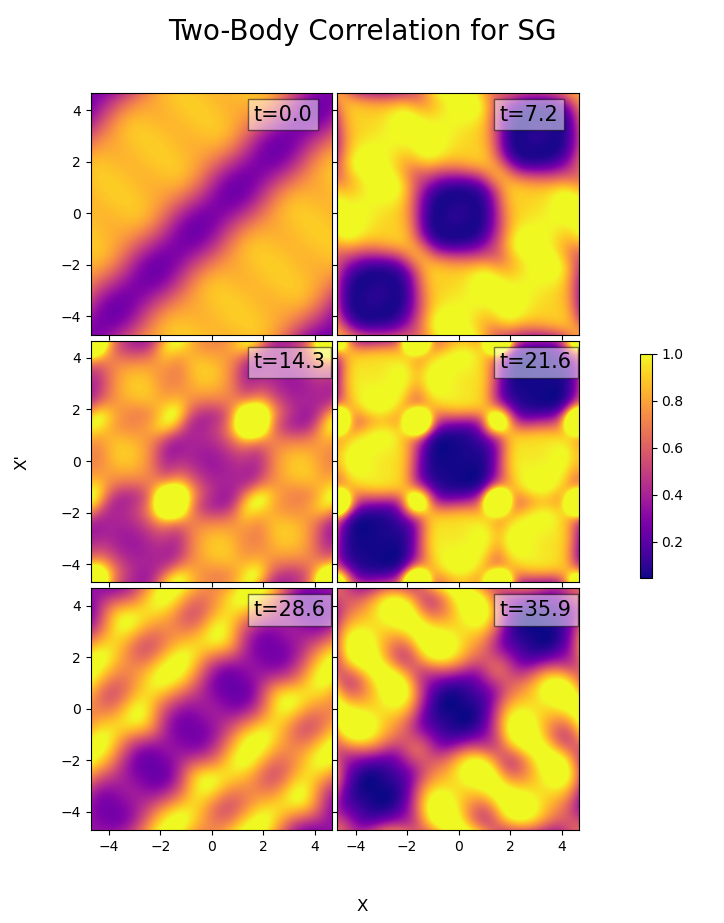}
    \caption{Second-order Glauber correlation ($g^{(2)}(x,x';t)$) at different times for the sine-Gordon dynamics, demonstrating oscillations between the fragmented superfluid and Mott-insulator phases. The distinct correlation holes along the diagonal signifies the Mott state, whereas the structureless diagonal extinction signifies fragmented superfluid phase. The system parameters and quench protocol are kept as the same as in Fig.~\ref{fig:sg_1B}. All quantities shown are dimensionless; refer to the text for further details.}
    \label{fig:sg_2B}
\end{figure}
\begin{figure}
    \centering
    \includegraphics[width=0.5\textwidth]{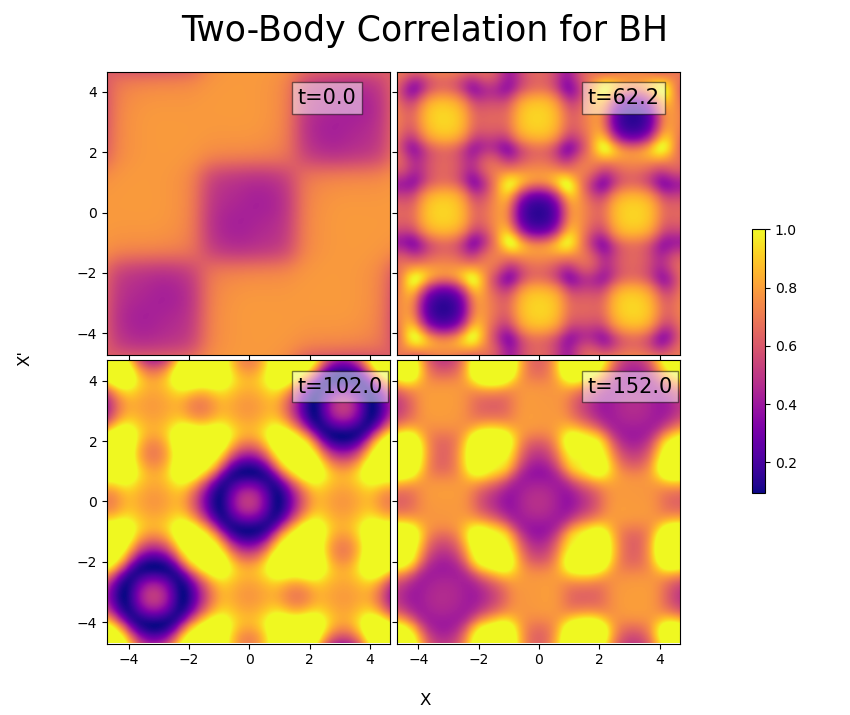}
    \caption{Second-order Glauber correlation ($g^{(2)}(x,x';t)$) at different times for the Bose-Hubbard dynamics, demonstrating oscillations between the non-fragmented superfluid and Mott-insulator phases. The distinct correlation holes along the diagonal signifies the Mott state, whereas the uniform two-body correlation across the lattice signifies the superfluid phase. The system parameters and quench protocol are kept as the same as in Fig.~\ref{fig:bh_1B}. All quantities shown are dimensionless; refer to the text for further details.}
    \label{fig:bh_2B}
\end{figure}

We calculate the two-body correlation function $g^{(2)}(x,x^{\prime};t)$ using Eq.(6) and plot them in Fig.~\ref{fig:sg_2B} and Fig.~\ref{fig:bh_2B} for the same selected times of first-order correlation dynamics. For a perfect SF phase, $g^{(2)}(x,x^{\prime})$ $\approx$ 1 for all $x$ and $x^{\prime}$--representing both the intra- as well as inter-well correlation. For a perfect Mott phase, $g^{(2)}(x,x^{\prime};t)$ exhibits correlation hole along the diagonal with vanishing correlation $g^{(2)}(x,x^{\prime})$ $\approx$ 0 and $g^{(2)}(x,x^{\prime})$ $\approx$ 1 for off-diagonal part ($x$ $\neq$ $x^{\prime}$).  Correlation hole is developed due to strong antibunching effect, two particles can never be simultaneously found in the same well due to strong repulsion. 

The two-body correlation dynamics for SG regime is presented in Fig.~\ref{fig:sg_2B}. At time $t=0$, we find extinction of diagonal correlation, but no distinct correlation hole is developed. It indeed signifies the strongly interacting fragmented SF phase. At $t=7.2$, when the system enters the Mott phase, clear structure of correlation hole is developed. The three completely extinguished Mott lobes along the diagonal signify three correlation holes. In the next slot of dynamics, the collapse-revival happen between the fragmented SF and Mott phases in the average time scale $\approx$ $7.0$.

The corresponding dynamics for BH is presented in Fig.~\ref{fig:bh_2B} for the same set of time slots as chosen for one-body correlation dynamics. At time $t=0$, the two-body correlation is maintained across the lattice with clear lattice structure along the diagonal, which signifies the initial state is a superfluid. At the time of first Mott entry $t_{entry}^{M1}$ = $62.2$, distinct correlation hole is built up which is further maintained till time $102.0$ with redistribution of diagonal and off-diagonal correlation. The SF phase is revived at time $152.0$ with distinct lattice structure along the diagonal and the off-diagonal correlation across the lattice.

\section{How to distinguish sine-Gordon from Bose-Hubbard dynamical regimes}
\label{sec:VI}

Here we prescribe four key measures: 
1) {\it {Time required for first Mott phase entry}} $t_{entry}^{M1}$;
2) {\it {Dynamical fragmentation}}; 
3) {\it{Many-body Information entropy}}; 
4) {\it{Two-body local correlation}} to distinguish sine-Gordon and Bose-Hubbard dynamics. 

\begin{figure}
    \centering
    \includegraphics[width=0.4\textwidth]{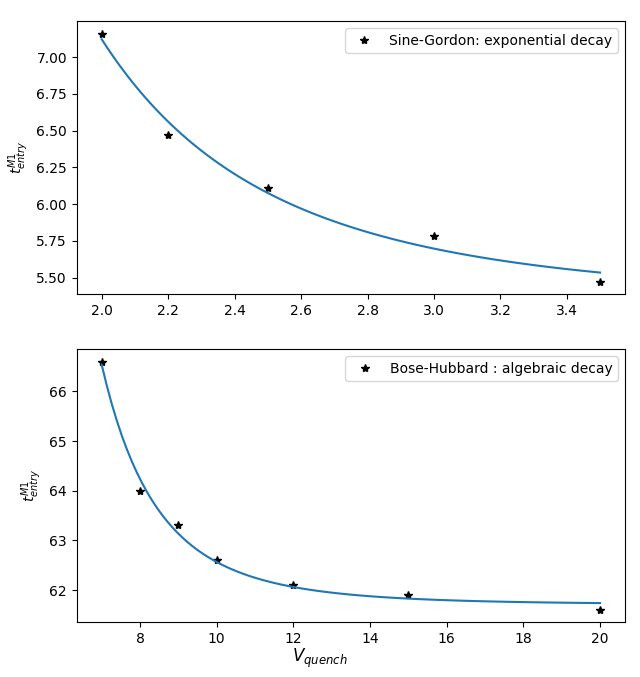}
    \caption{Plot of the time for first Mott entry, $t^{MI}_{entry}$ for various lattice depth. The top panel corresponds to the SG dynamics, while the bottom panel shows the corresponding plot for the BH dynamics. Notably, the entry time $t^{MI}_{entry}$ follows an algebraic law during the BH dynamics, whereas an exponential decay is observed in the SG dynamics. All quantities are dimensionless. }
    \label{fig:mott_entry}
\end{figure}

In Fig.~\ref{fig:mott_entry}, we plot $t_{entry}^{M1}$ for sine-Gordon dynamics (upper panel) and the same for Bose-Hubbard dynamics (lower panel) for various quench parameters. For SG transition we scan only the pinning zone, lattice depth is varied between 2.0 and 3.5 keeping the interaction stregth $\lambda$ fixed to 1.01. With larger lattice depth,  $t_{entry}^{M1}$ reduces which can be fitted to an exponential decay. Whereas for BH dynamics we choose large lattice depth quench between $7.0$ to $20.0$. For all quench processes the initial states remain the same as described earlier. The required time for Mott entry is very high and slowly falls with higher lattice depth quench, we observe the algebraic decay as observed in 
experiment. 

Thus the dynamics for sine-Gordon regime is quite fast compared to Bose-Hubbard time scale. The possible explanation is the initial state for SG dynamics is already correlated and fragmented and carries off-diagonal correlation with time, in contrast, the initial state for BH dynamics is superfluid with complete correlation across the lattice. 
\begin{figure}
    \centering
    \includegraphics[width=0.4\textwidth]{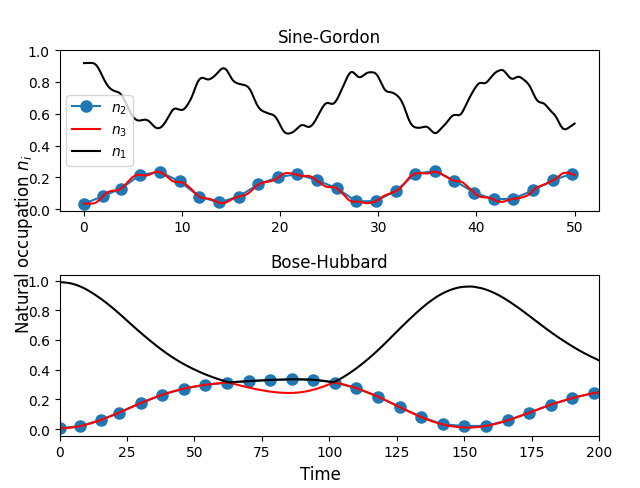}
    \caption{Natural occupation as a function of time for three significantly occupied orbitals. The upper panel corresponds to sine-Gordon dynamics and the lower panel corresponds to Bose-Hubbard dynamics. See the text for details.}
    \label{fig:nopr}
\end{figure}

In Fig.~\ref{fig:nopr}, we present the signature of dynamical fragmentation which depict the natural orbital occupation. Fragmentation means when the macroscopic occupation in several single-particle functions is energetically more favourable than the accumulation of all bosons in a single orbital. Dynamical fragmentation quantifies how the fragmentation changes during dynamical evolution. Although both in SG and BH dynamics the post quench state is a Mott state, we find stringent difference in terms of fragmentation as depicted in Fig.~\ref{fig:nopr}. Here we plot the occupation in lowest three orbitals $n_i(t)\ (i=1,2,3)$ for the same quench parameters as in section III. For BH regime, the initial state is purely condensate $\left[n_1(t=0)=1, n_2(t=0)=0,n_3(t=0)=0\right]$. The initial state configuration is simply $\vert 1,0,0 \rangle$ and the many-body state should be described by the Gross-Pitaevskii mean-field state. 
With time, second and third orbitals start to contribute and exactly at the time of $t_{entry}^{M1}$ it becomes threefold fragmented; occupation in each orbital becomes 33 \% at time $t=62.2$.  
The many-body wavefunction can now be represented by $\vert 1,1,1 \rangle$, the Mott phase can be renamed as fragmented Mott. With further increase in time $n_1$ and $n_2$ overlap and $n_3$ lowers and at a later time the three-fold fragmentation is reached again at time $t=102$. Thus between $t$= 62.2 and $t$= 102, the occupation in the lowest three orbitals reshuffle and we can take it as holding time for Mott state. The superfluid state revives at time $t= 152$.  The superfluid - Mott collapse revival dynamics can now be described transition between non-fragmented to three-fold fragmented many-body states. 

In contrast, in the sine-Gordon regime (upper panel Fig.~\ref{fig:nopr}) clearly shows the initial state is fragmented $\left[n_1(t=0)=0.92, n_2(t=0)=0.036, n_3(t=0)=0.036\right]$ with contribution of $n_4$ = $0.006$ and $n_5$ is $0.002$ (not shown in the graph). 
With time fragmentation is built up, population in first orbital decreases and population in second and third orbitals increase but at time $t_{entry}^{M1}$, three-fold fragmentation is not reached; although collapse revival scenario is seen at a later time. At time $t_{entry}^{M1}$ the state is a Mott state with clear diagonal correlation and extinction of off-diagonal correlation; but the many-body state is not three-fold fragmented. Thus the post quench state in sine-Gordon dynamics is a many-body fragmented but not the $\frac{1}{N}$ fragmented as observed in Bose-Hubbard dynamics.

\begin{figure}
    \centering    \includegraphics[width=0.45\textwidth]{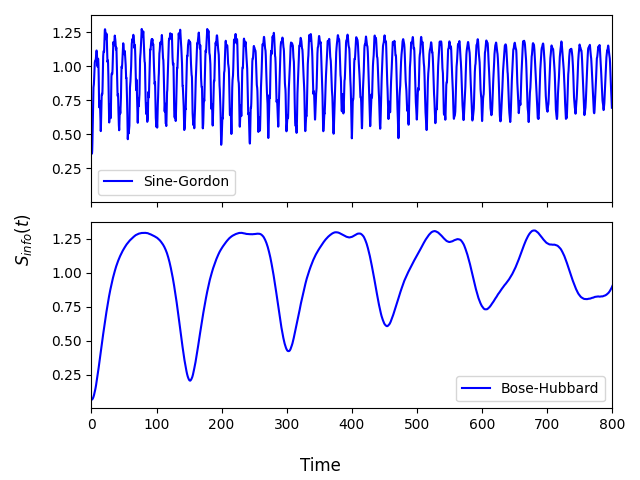}
    \caption{ The Shannon information entropy (SIE) is plotted as a function of time. In the top panel, the SIE for the sine-Gordon (SG) dynamics is shown. The non-zero initial entropy at $t=0$ indicates that the initial state is fragmented rather than fully coherent and it exhibits very fast oscillations with time. In the bottom panel, the SIE for the Bose-Hubbard (BH) dynamics is displayed. The initial entropy being close to zero implies that the state is a pure superfluid, after exhibiting few collapse-revival scenario with a long period of time it approaches relaxation. }
    \label{fig:entropy}
\end{figure}

 In Fig.~\ref{fig:entropy}, we plot $S_{info}(t)$ in the long time dynamics. We find modulated oscillation in BH dynamics and unmodulated oscillation in SG dynamics. For BH dynamics initial entropy is close to zero as one single orbital contributes; associated oscillation exhibit the collapse revival dynamics but not with definite time scale. Then the possibility of complete revival gradually decreases  which signifies possible relaxation at much longer time. In contrast, in SG dynamics, the unmodulated oscillation in the entropy evolution at a very fast rate and with a definite period even for quite long time can not predict any possible relaxation. We do not conclude when and how relaxation will happen in sine-Gordon dynamics. The system may eventually relax at very long time which is beyond the scope of present simulation. 

 To obtain the GOE estimate for the information entropy, we set $n_i$= $\frac{N}{M}$ for all $i$ in Eq.(7). We obtain $S_{info}^{GOE}$ = $-\sum_{i=1}^{M} \frac{1}{M} \ln(\frac{1}{M}) = \ln(M)$. For $N=3$ bosons in $S=3$ sites as the post quench state is a three-fold fragmented, three natural orbital exhibits equal population each of $33 \%$. Thus in GOE estimate we put $M=3$ which results to $S_{info}^{GOE}= 1.09$. Fig.~\ref{fig:entropy} exhibits that in the BH dynamics the maximum entropy in the relaxed state is approaching GOE prediction and in the SG dynamics, entropy oscillates about the GOE value.  

\begin{figure}[b]
    \centering
    \includegraphics[width=0.45\textwidth]{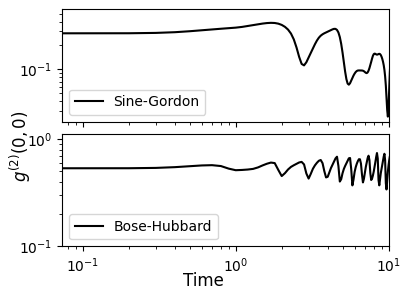}
    \caption{Local two-particle correlation function $g^{(2)}(0, 0)$ as a function of time for sine-Gordon (top) and Bose-Hubbard dynamics (bottom). All quantities are dimensionless. See the text for details. }
    \label{fig:g200}
\end{figure}
In Fig.~\ref{fig:g200}, we show the local two-particle correlation $g^{(2)}(0,0)$ as a function of time for the lattice depth quench parameter as in Fig.~\ref{fig:sg_2B} and Fig.~\ref{fig:bh_2B}. In both SG and BH dynamics, initially $g^{(2)}(0,0)$ remains very small.  Specifically, in SG dynamics, with time $g^{(2)}(0,0)$ slightly rises up with time and then exhibits peculiar oscillatory behavior with large modulation depth. Whereas in BH dynamics, $g^{(2)}(0,0)$ remains constant for significant time and then exhibits some oscillation. 

\section{Conclusions}
\label{sec:VII}
In conclusion, we have applied a numerically exact many-body approach, MCTDHB, to study the superfluid to Mott-insulator transition for two different quench protocol. The quench dynamics is studied by the time evolution of one-body and two-body correlation functions. We study sine-Gordon quench dynamics when the strongly interacting superfluid in the vanishing lattice is suddenly perturbed by a weak lattice that mimics the sine-Gordon phase transition. In the other quench protocol, we study the dynamics when weakly interacting superfluid is suddenly perturbed by a deep lattice that mimics the Bose-Hubbard phase transition. For the first case, the dynamical features are dictated by the strong initial correlation, whereas in the second case, the dynamical features are controlled by deep lattice. The two kinds of dynamics finally lead to Mott localization, but in different time scale. The intermediate dynamical features in correlation measures clearly distinguish the two kinds of dynamics which happen in different time scales. The dynamical fragmentation and the time for first Mott entry are established as two metrics to distinguish the SG and BH dynamics. The additional calculation of information entropy exhibits clear relaxation process for BH dynamics, whereas in SG dynamics it is hard to access when and how relaxation will happen. 

We conclude by observing that the differences between BH and SG dynamical regimes observed in this work 
and their characterization can be concretely studied in feasible experiments in future. We further envision that similar approaches could be employed for long-range dipolar interaction. Our method solves the many-body Schr\"odinger equation in the continuum and thus capture the physics at play in the actual quantum simulators and not in the simulated lattice models.  

\section*{Acknowledgements}
B C would like to pay gratitude to Grigory E. Astrakharchik for providing the data of exact diagonalization of Bose-Hubbard model which has been used in Fig.1. A G acknowledges CNPq-Conselho Nacional de Desenvolvimento Científico e Tecnológico (Brazil) grant no. 306209/2022-0. B C acknowledges FAPESP grant Process No. 2023/06550-4.

\appendix
\section{Results for different system size}
In this Appendix, we show the Fig.~\ref{fig:entropy} result of the main text for larger lattice size with unit filling factor one to demonstrate that the results presented in the main text can be generalized. Similar to the main text results for $N=3$ bosons in $S=3$ lattice sites, we present the information entropy measure with $N=5$ bosons in $S=5$ sites; $N=7$ bosons in $S=7$ sites. We keep the parameters in quench protocol as given in the main text. To obtain the proper convergence we used $M=8$ orbitals for $N=5$ bosons in $S=5$ lattice sites and $M=10$ orbitals for $N=7$ bosons in $S=7$ lattice sites. In both cases we observe the same physics as concluded in the main text. In the SG dynamics, we observe vary fast oscillation which mimics that collapse-revival happens in a very short time scale and we do not observe any possibility of relaxation in the present time scale. In the BH dynamics, collapse-revival happens in  large time scale and signature of relaxation is clearly exhibited. 

\begin{figure}
    \centering    \includegraphics[width=0.45\textwidth]{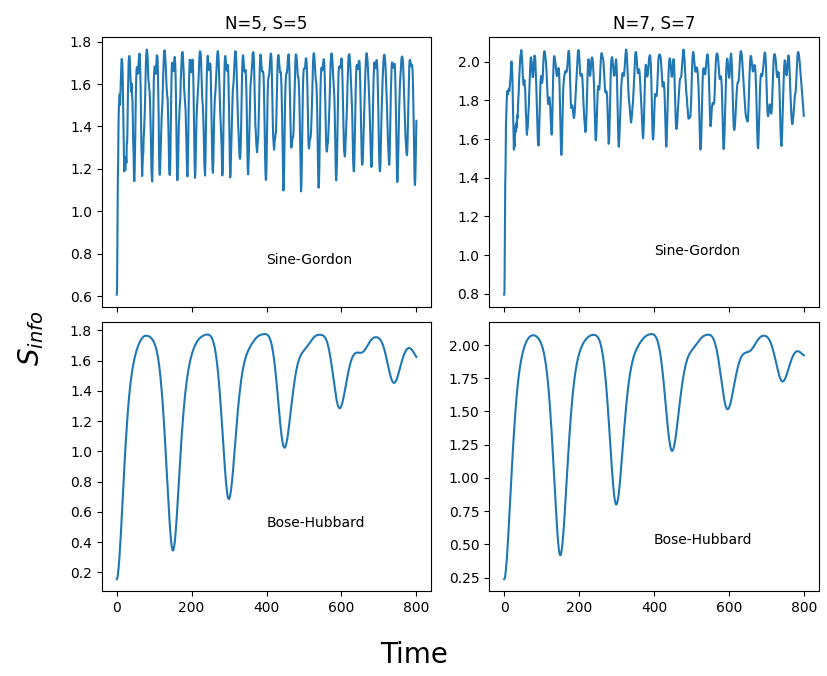}
    \caption{ The Shannon information entropy (SIE) is plotted as a function of time. The left column is for $N=5$ bosons in $S=5$ lattice sites. The top panel is for SG dynamics which exhibits very fast oscillation and the lower panel is for BH dynamics which shows a possibility of relaxation. The Right column explains the same physics for $N=7$ bosons in $S=7$ lattice sites. All quantities are dimensionless. See the text for further details.}
    \label{fig:entropy_all}
\end{figure}

The GOE predicted value for $N=5$ in $S=5$ sites would be $S_{info}^{GOE}=1.61$ as the system now becomes fifth-fold fragmented. Left column of Fig.~\ref{fig:entropy_all} exhibits that in the BH dynamics, the maximum entropy is reaching the GOE estimate and in the SG dynamics, entropy oscillates about the GOE value. For $N=7$ bosons in $S=7$ sites, the system is seventh-fold fragmented and the GOE estimate is $S_{info}^{GOE}=1.94$. 
Right column of Fig~\ref{fig:entropy_all}, exhibits that when entropy oscillates about this GOE value in SG dynamics, the post quench state in BH dynamics try to relax to the GOE estimated maximum entropy state. 

\section{System parameters}
In this appendix we discuss the parameters for the numerical simulations in the main text. We perform the simulations of $N=3$ bosons with $M=3$ orbitals.
The periodic lattice of lattice depth $V$ and wave vector $k$ is parameterized as 
\begin{equation}
V(x) = V\sin^2(kx)
\end{equation}
where $k$ is the wave vector. We choose the wavelength $\lambda_0 \simeq 532.2 nm$  which is compatible with the experiments in ultracold atomic gases. It gives the wave vector $k \simeq 5.913 \times 10^{6}$ m$^{-1}$. 

\subsection{Lengths}
In MCTDH-X simulations, we choose to set the unit of length $\bar{L} \equiv \frac{\lambda_0}{3}= 177.33$ nm, which makes the minima of the primary lattice appear at integer values in dimensionless units, while the maxima are located at half integer values. $x=0$ is the center of the lattice which  can host an odd number of lattice sites $S$. In our numerical simulation, we consider an integer filling of $N=3$ bosons in $S=3$ sites and run simulations with 512 grid points. \\

\subsection{Energies}

The unit of energy $\bar{E}$ is defined in terms of the recoil energy of the primary lattice, i.e. $E_r \equiv \frac{\hbar^2 k^2}{2m} \simeq  1.211 \times 10^{-30}$ J  with $m$ $\simeq$ 86.909 Da, the mass of $^{87}$Rb atoms.  Thus we define the unit of energy as $\bar{E} \equiv \frac{ \hbar^2} {2m L^2}$ = $ E_r (\frac{3}{\pi})^2$ = $ 1.104 \times 10^{-30}$ J. In typical experiments in deep periodic optical lattice, the depth is varied in the around few tens of  recoil energies. In our simulations, we probe similar regimes: $V$ $\in$ $\left[ 8 E_r, 20 E_r\right]$. In typical experiments in shallow lattice, the depth is varied between few recoil energy. In our simulation we probe similar regimes: $V$ $\in$ $\left[ 1.5 E_r, 3.5 E_r \right]$. The on-site interactions are kept fixed $\lambda=0.05 E_r$ for weakly interacting superfluid  and $\lambda= 1.01 E_r$ for strongly interacting superfluid, these values can be achieved in the ultracold quantum simulators.

\subsection{Time}
The unit of time is defined from the unit of length as $\bar{t}$  $\equiv \frac{ m \bar{L}^2}{\hbar}$ = $5.051 \mu$s. For both the dynamics we probe up to $800\bar{t}$ which is $\sim 4.04$ms.

\end{document}